\documentclass[a4paper]{jpconf}

\usepackage{graphicx}

\begin{document}

\begin{flushright}
{\large \tt TTK-11-45}
\end{flushright}

\title{Bayes and present dark matter direct search status}

\author{Chiara Arina}

\address{Institut f\"ur Theoretische Teilchenphysik und Kosmologie, RWTH Aachen, 52056 Aachen, Germany}

\ead{chiara.arina@physik.rwth-aachen.de}

\begin{abstract}
Recently there has been a huge activity in the dark matter direct detection field, with the report of an excess from CoGeNT 
and CRESST along with the annual modulated signal of DAMA/Libra and the strong exclusion bound 
from XENON100. We analyse these results within the framework of Bayesian inference and evidence. Indeed bayesian methods are well suited for marginalizing over experimental systematics and background. We present the results for 
spin-independent interaction on nucleus with particular attention to the low dark matter mass region and the compatibility between experiments. In the same vein we also investigate the impact of astrophysical uncertainties on the WIMP preferred parameter space within the class of isotropic dark matter velocity distributions.
\end{abstract}

\section{Introduction}
Dark Matter (DM) direct searches experience a fervent activity with the report of an annual modulation by the CoGeNT collaboration~\cite{Aalseth:2011wp} along with the established DAMA/Libra (DAMA hereafter) signal~\cite{Bernabei:2008yi}. The exclusion bound of the XENON100 collaboration~\cite{:2011hi} (Xe100 hereafter), obtained via a profile likelihood approach~\cite{Aprile:2011hx}, puts to test those excesses. The aim of this proceeding is to update the bayesian analysis of~\cite{Arina:2011si} with the CoGeNT data 2011 and to investigate the mutual agreement/disagreement between those experiments for spin-independent interaction. As a novelty, we present the assessment of the need of including astrophysical uncertainties in the direct detection (DD) analysis, with the use of the Bayesian evidence.

\section{Bayesian framework and experimental likelihoods}
The bayesian statistics is a well defined framework that allows the inclusion of experimental systematics and backgrounds, as well as astrophysical uncertainties common to all experiments. Indeed the posterior probability density function (pdf) is given by Bayes' theorem:
\begin{equation}
\label{eq:BT}
\mathcal{P}( \theta \mid X) = \pi(\theta)\ \frac{{\mathcal L}(X | \theta)}{\mathcal{Z}(X)}\,,
\end{equation}
where the likelihood $\mathcal{L}$ describes how the theoretical model, with free parameters $\theta$, describes the data $X$. The prior probability density $\pi$ encodes the state of knowledge on $\theta$ before observing the data and is therefore independent of $X$. The posterior pdf is sampled via MCMC techniques using the package \texttt{CosmoMC}~\cite{Lewis:2002ah,cosmomc_notes} and then marginalized over nuisance parameters. For DD purposes, the important quantity is $\mathcal{P}_{\rm marg}(m_{\rm DM},\sigma_n^{SI})$, where $m_{\rm DM}$ is the DM mass and $\sigma_n^{SI}$ is the DM-nucleus cross-section. 

We take as nuisance parameters the mean velocity of the DM at the sun position $v_0$, the escape velocity from the DM halo $v_{\rm esc}$ and the DM density at the sun position $\rho_{\odot}$ (altogether named astrophysical variables). We consider two models for the DM velocity distribution: SMH (Standard Model Halo) -- that is a simple Maxwellian halo with astrophysical variables fixed at their mean value -- and NFW -- namely we derive the velocity distribution from this DM density profile and marginalize over all possible values of $v_0$, $v_{\rm esc}$ and $\rho_\odot$ as well as profile parameters. For all technical details about the analysis, prior choices and likelihood definitions, we refer to~\cite{Arina:2011si}. Here we just recall that the main systematic for Xe100 is the scintillation efficiency, while for DAMA the quenching factor on Sodium $q_{\rm Na}$ and on Iodine $q_{\rm I}$ are marginalized over all their experimental range. Regarding the new CoGeNT data, the likelihood is given by the product of two gaussian distributions, one for the total rate and one for the modulated signal. The data binning, analysis and subtraction of cosmogenic peaks follow closely~\cite{Aalseth:2011wp} and~\cite{collar}. A flat and exponential background, which does not modulate in time and which is described by 3 nuisance parameters, is added on top of the DM signal. The exponential background accounts for the effect of a bad rejection of surface events near threshold.

In Eq.~\ref{eq:BT} the denominator $\mathcal{Z}(X)$, called Bayesian evidence, is defined as the average of the likelihood over the prior for a specific model $\mathcal{M}$
\begin{equation}
\label{eq:evidence}
\mathcal{Z} = \int \mathcal{L}(X|\theta) \pi(\theta) d^D \theta\,,
\end{equation}
where $D$ is the dimensionality of the parameter space. This quantity is of crucial importance in model comparison, which is defined as the ratio of posterior probabilities. A model $\mathcal{M}_0$ can be compared with $\mathcal{M}_1$, which has extra parameters, through the Bayes factor $B$, see {\it e.g.}~\cite{Kunz:2006mc,Trotta:2008qt}:
\begin{equation}
\frac{\mathcal{P}(\mathcal{M}_0|X)}{\mathcal{P}(\mathcal{M}_1|X)}   =  \frac{\mathcal{Z}_0}{\mathcal{Z}_1}\frac{\pi(\mathcal{M}_0) }{\pi(\mathcal{M}_1) }  = B_{01}\frac{\pi(\mathcal{M}_0) }{\pi(\mathcal{M}_1) }  \,,
\end{equation}
where $\pi(\mathcal{M}_i)$ is the prior probability of each model. The Bayes factor automatically favours simpler models unless the data justify the complexity of more complicated alternatives, because of the marginalization procedure used to calculate the evidence, Eq.~\ref{eq:evidence}. In the following we will compare the model SMH with NFW, the evidence for each model being computed with \texttt{MULTINEST}~\cite{Feroz:2007kg,Feroz:2008xx}.

\begin{figure}[t!]
\begin{minipage}[t]{0.5\textwidth}
\centering
\includegraphics[width=1.1\columnwidth]{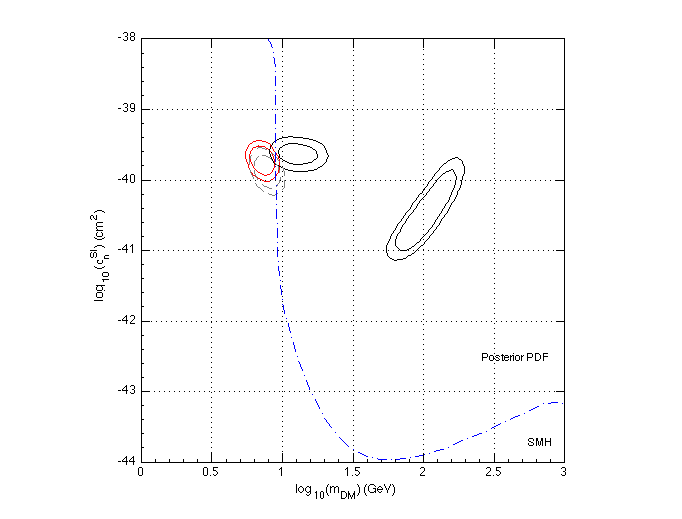}
\end{minipage}
\begin{minipage}[t]{0.5\textwidth}
\centering
\includegraphics[width=1.1\columnwidth]{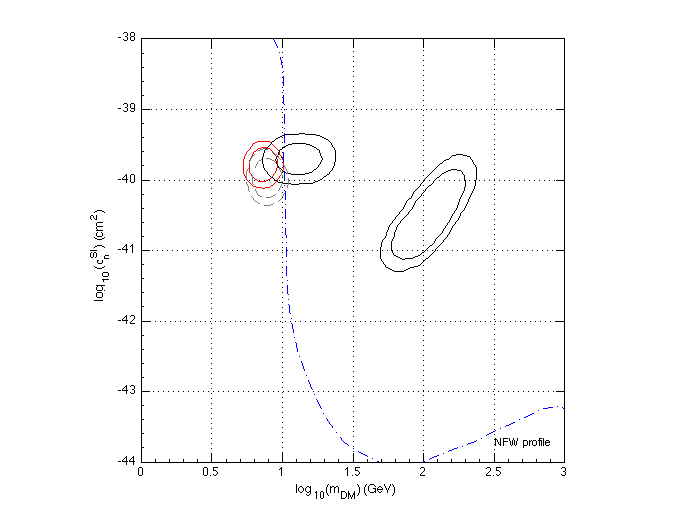}
\end{minipage}
\caption{{\it Left}: 2D credible regions for the individual experimental bounds and regions assuming the SMH, combined in a single plot. For DAMA (black solid), CoGeNT (gray dashed) and the combined fit (red solid) we show the 90\% and 99\% contours.
The blue dashed line represents the $90_S$\% confidence level (C.L.) for Xe100. {\it Right}: Same as left for the NFW model.
\label{fig1}}
\end{figure}

\section{Results}
Figure~\ref{fig1} summarizes the 2D inference in the $\{m_{\rm DM}, \sigma_n^{SI}\}$ plane for individual experiments (blue dashed curve for Xe100, gray dashed and solid black contours for CoGeNT and DAMA respectively) and for the combined fit (solid red lines), in the SMH and NFW model (left and right panel respectively). Table~\ref{tab1} resumes the preferred values for the DM and astrophysical observables in each model (recalling that $v_0$, $v_{\rm esc}$ and $\rho_\odot$ are fixed for SMH model). Comparing the left and right panel one notes that the closed regions become wider, while the Xe100 exclusion bound moves slightly to the right, improuving the agreement between various search results. This is a volume effect due to the marginalization procedure over the astrophysical variables. In both models, the inclusion of uncertainties on the scintillation energy leads to compatibility between CoGeNT, the combined fit and Xe100 exclusion bounds at $90_S$\% C.L., while leaving a marginal compatibility with DAMA at 99\% C.L.. It is always possible to combine experiments and find a common region, but actually the question is whether the preferred point for $\{m_{\rm DM}, \sigma_n^{SI}\}$ is a good fit for both experiments and what are the values for the nuisance parameters that are selected. Notice from table~\ref{tab1} that the combined fit chooses values which are in line with the one of CoGeNT and DAMA alone, within the statistical errors. The large deviation from DAMA alone fit concerns the quenching factor on Sodium: as it has been shown in figure 2 of~\cite{Arina:2011si} the 1D posterior pdf for $q_{\rm Na}$ is flat all along its prior range, while the combined fit is strongly peaked at the value $q_{\rm Na} = 0.57 \pm 0.03$. This can be understood thinking that the larger $q_{\rm Na}$ the lower the DM mass should be to account for the DM modulated signal. The selected value for the combined fit therefore pushes the DAMA region towards the direction of the CoGeNT region.

Table~\ref{tab2} resumes the results for model comparison in the case of individual experiments and for the DAMA+CoGeNT combined fit. A positive (negative) value for $ \ln B$ represents an increase (decrease) of the support in favour of the simplest model given the observed data. The strength of evidence is given by the empirical `Jeffreys scale' (see table 1 in~\cite{Trotta:2008qt}), where threshold values are set for inconclusive, moderate or strong evidence: {\it e.g.} $\ln B=5$ states that the odds against the most complicated models are 150:1 and corresponds to a probability of 0.993. This holds in the case where the probability of each model is the same, namely $\pi(\mathcal{M}_0)= \pi(\mathcal{M}_1) = 1/2$. Looking at $\ln B$ in table~\ref{tab2} it is clear that there is strong or moderate evidence against NFW model for single experiments. Therefore data at present time do not allow a determination of astrophysical observables. On the contrary there is a very strong evidence for NFW for the combined fit, namely experiments need to adjust astrophysical observable values to find a common agreement. Similar results hold for the whole class of spherically symmetric DM halos.

\begin{table}[t!]
\small
\caption{1D posterior pdf modes and $90\%$ credible intervals for the circular velocity $v_0$,
escape velocity $v_{\rm esc}$, and the local DM density $\rho_{\odot}$ for the 2 DM halo model considered in this work. \label{tab1}}
\begin{center}
\lineup
\begin{tabular}{ ll|ccccc }
\br
Model & & $m_{\rm DM}$ (GeV) & $\sigma_n^{SI} ({\rm cm^2})$ & $v_0$ (km/s) & $v_{\rm esc}$ (km/s) & $\rho_{\odot}$  ($ {\rm GeV/ cm^3}$) \\ 
\br
\bfseries{SMH}& DAMA& 12 & $2.2 \times 10^{-40}$  & 230 & 544 &0.4 \\
 & CoGeNT & 7.5 & $1.9 \times 10^{-40}$ &  230 &  544 &0.4  \\
& Combined & 6.9 &$1.8 \times 10^{-40}$ & 230 &544  & 0.4 \\
& Xenon100 &-- &-- &230 &544 & 0.4 \\
\mr
\bfseries{NFW} & DAMA & 12 & $1.7 \times 10^{-40}$ & $220^{+40}_{-20}$ & $558_{-16}^{+19}$ & $0.39_{-0.09}^{+0.15}$ \\
& CoGeNT & 7.4 & $9.7 \times 10^{-41}$ & $219_{-22}^{+39}$ & $557_{-15}^{+18} $ & $0.37_{-0.10}^{+0.15}$ \\
& Combined & 7. & $1.5 \times 10^{-40}$  & $214_{-21}^{+32}$ & $556_{-15}^{+14} $ & $0.35_{-0.09}^{+0.14}$ \\
& Xenon100 & -- & -- & $219_{-20}^{+43}$ & $559\pm 18$ & $0.37_{-0.08}^{+0.16}$  \\
\br
\end{tabular}
\end{center}
\end{table}

\begin{table}[t!]
\small
\caption{The odds against the NFW model for individual and combined experiments. \label{tab2}}
\begin{center}
\lineup
\begin{tabular}{ l|ccc }
\br
& DAMA & CoGeNT & Combined  \\ 
\br
$\ln B$ & 5.3 & 3.9 & -32\\
\br
\end{tabular}
\end{center}
\end{table}

\section{Conclusion}

We have employed Bayesian inference to study the concordance between various DD experiments and model comparison to infer the necessity of a more complicated model than the SMH at present time. The outcomes of the analysis may be summarized in few main points:
\begin{itemize}
	\item the combined fit of DAMA and CoGeNT requires parameters closely in line with those of individual fits, the only exception being a large quenching factor on Sodium;
	\item the CoGeNT region and the combined fit are compatible at $90_S$\% C.L. with the Xe100 exclusion bounds, while DAMA is only marginally allowed at 99\% C.L.;
	\item a combination of at least two experiments demonstrates the ability of constraining astrophysical variables, on the contrary of a single DD experiment.
\end{itemize} 

Remarkably the excess region falls in the same ballpark of $\{m_{\rm DM}, \sigma_n^{SI}\}$ values that could explain the excess of events reported by the CRESST experiment~\cite{Angloher:2011uu}. The CoGeNT collaboration has presented an ongoing analysis on the rejection of surface events near threshold~\cite{collarTaup}, which could account for at least 50\% of the counts at low energies: the total rate would therefore be accommodated by smaller DM cross-sections than those in figure~\ref{fig1}.

\subsection*{Acknowledgments}
It's a pleasure to thank R. Trotta for discussions on model comparison and comments on the manuscript and the CoGeNT collaboration for making data available for public analysis. This work acknowledges use of the COSMO computing resource at CP3 of Louvain University. Published under licence in {\it Journal of Physics: Conference Series} by IOP Publishing Ltd.

\section*{References}
\bibliographystyle{iopart-num.bst}
\bibliography{biblio}

\providecommand{\newblock}{}
\begin{thebibliography}{10}
\expandafter\ifx\csname url\endcsname\relax
  \def\url#1{{\tt #1}}\fi
\expandafter\ifx\csname urlprefix\endcsname\relax\def\urlprefix{URL }\fi
\providecommand{\eprint}[2][]{\url{#2}}

\bibitem{Aalseth:2011wp}
Aalseth C, Barbeau P, Colaresi J, Collar J, Leon J {\em et~al.\/} 2011
  (\textit{Preprint} \eprint{1106.0650})

\bibitem{Bernabei:2008yi}
Bernabei R {\em et~al.\/} (DAMA) 2008 {\em Eur. Phys. J.\/} {\bf C56} 333--355
  (\textit{Preprint} \eprint{arXiv:0804.2741})

\bibitem{:2011hi}
Aprile E {\em et~al.\/} (XENON100) 2011  (\textit{Preprint} \eprint{1104.2549})

\bibitem{Aprile:2011hx}
Aprile E {\em et~al.\/} (XENON100 Collaboration) 2011  (\textit{Preprint}
  \eprint{1103.0303})

\bibitem{Arina:2011si}
Arina C, Hamann J and Wong Y~Y 2011 {\em JCAP\/} {\bf 1109} 022
  (\textit{Preprint} \eprint{1105.5121})

\bibitem{Lewis:2002ah}
Lewis A and Bridle S 2002 {\em Phys. Rev.\/} {\bf D66} 103511
  (\textit{Preprint} \eprint{astro-ph/0205436})

\bibitem{cosmomc_notes}
Lewis A and Bridle S {CosmoMC Notes} http://cosmologist.info/notes/CosmoMC.pdf

\bibitem{collar}
Collar J {\em How to use time-stamped CoGeNT data.\/}

\bibitem{Kunz:2006mc}
Kunz M, Trotta R and Parkinson D 2006 {\em Phys.Rev.\/} {\bf D74} 023503
  (\textit{Preprint} \eprint{astro-ph/0602378})

\bibitem{Trotta:2008qt}
Trotta R 2008 {\em Contemp.Phys.\/} {\bf 49} 71--104 (\textit{Preprint}
  \eprint{0803.4089})

\bibitem{Feroz:2007kg}
Feroz F and Hobson M 2007  (\textit{Preprint} \eprint{0704.3704})

\bibitem{Feroz:2008xx}
Feroz F, Hobson M and Bridges M 2008  (\textit{Preprint} \eprint{0809.3437})

\bibitem{Angloher:2011uu}
Angloher G, Bauer M, Bavykina I, Bento A, Bucci C {\em et~al.\/} 2011
  (\textit{Preprint} \eprint{1109.0702})

\bibitem{collarTaup}
Collar J {\em talk @ TAUP 2011 conference\/}

\end{thebibliography}

\end{document}